# An effective magnetic field from optically driven phonons

T. F. Nova[1] *, A. Cartella[1], A. Cantaluppi[1], M. Först[1], D. Bossini[2] #, R. V. Mikhaylovskiy[2],

A.V. Kimel[2], R. Merlin[3] and A. Cavalleri[1, 4] *

[1]Max Planck Institute for the Structure and Dynamics of Matter, 22761 Hamburg, Germany

[2]Radboud University, Institute for Molecules and Materials, 6525 AJ Nijmegen, The Netherlands

[3]Department of Physics, University of Michigan, Ann Arbor, Michigan 48109-1040, USA

[4]University of Oxford, Clarendon Laboratory, Oxford OX1 3PU, UK

*Corresponding authors

[#] Present address: The University of Tokyo, Institute for Photon Science and Technology,

7-3-1 Hongo, Bunkyo-ku, Tokyo 113-0033, JAPAN

**Light fields at THz and mid-infrared frequencies allow for the direct excitation of collective modes in condensed matter, which can be driven to large amplitudes. For example, excitation of the crystal lattice [1,2] has been shown to stimulate insulator-metal transitions [3,4], melt magnetic order[5,6] or enhance superconductivity[7,8,9]. Here, we generalize these ideas and explore the simultaneous excitation of more than one lattice mode, which are driven with controlled relative phases. This nonlinear mode mixing drives rotations as well as displacements of the crystal-field atoms, mimicking the application of a magnetic field and resulting in the excitation of spin precession in the rare-earth orthoferrite $ErFeO_3$. Coherent control of lattice rotations may become applicable to other interesting problems in materials research, for example as a way to affect the topology of electronic phases.**

ErFeO$_3$ is an antiferromagnetic insulator that crystallizes in an orthorombically distorted perovskite structure, as shown in Fig. 1a (space group: Pbnm). Because of the Dzyaloshinskii-Moryia interaction, the spins are canted and result in a small ferromagnetic moment (saturated by a 7.5 mT field[10]) along the *c* axis (Fig. 1b).

In our experiments, femtosecond mid-IR pulses at 20 THz frequency were tuned to drive the in-plane $B_{ua}$ and $B_{ub}$ phonons[11]. The eigenvectors of these modes are shown in Fig. 1c. The pump was linearly polarized and aligned at variable angles with respect to the *a* and *b* crystallographic axes. Due to the orthorhombic distortion in ErFeO$_3$, the infrared-active modes along the two axes exhibit different eigenfrequencies and oscillator strengths. As a consequence, even when the modes are excited simultaneously (e.g. with the light polarization at 45 degrees angle from either crystal axes), the two modes start with a nonzero relative phase, and advance in time at different rates. Hence, the total lattice polarization inside the material is elliptical (Fig. 1d). This is interesting, as coherent atomic loops[12] can break time reversal symmetry and stimulate a new class of opto-magnetic phenomena, which are independent of equilibrium multiferroicity[13,14,15].

These considerations are validated here by mid-infrared pump, optical probe experiments performed at 100 K. The Faraday rotation of a linearly polarized near-infrared pulse (800-nm wavelength) was measured as a function of time delay after direct lattice excitation. As shown in Figure 2a, the polarization of the probe was found to oscillate in time, revealing the coherent excitation of a number of Raman active modes (Fig. 2b). These include Raman phonons of $A_{1g} + B_{1g}$ and $B_{1g}$ symmetry (3.36 THz and 4.85 THz, respectively).

Strikingly, we also observed the excitation of a coherent quasi-antiferromagnetic magnetic mode (q-AFM, 0.75 THz) [16], associated with a modulation of the ferromagnetic moment along the *c* axis (Fig. 2c)[17].

Figure 3a displays the pump-wavelength-dependent amplitude of this mode, plotted with the reflectivity of the material and the real part of the dielectric function. The magnitude of the magnetic mode was larger for wavelengths near the zero crossing of the dielectric function (19.6 THz). The corresponding lattice-induced magnetization was estimated by normalizing the raw data against the penetration-depth mismatch between the pump and the probe (Fig. 3b, see Supplementary Section 6). The maximum precession amplitude was approximately 1.5% of the saturation magnetization. Also, as displayed in Fig. 3c, this amplitude scaled quadratically with the electric field of the pump, indicating a response proportional to the product of two phonon coordinates.

The coherent magnon oscillations were only observed for pump pulses polarized at +45 or − 45 degrees, in between the crystallographic *a* and *b* axes (Fig. 3d, red and orange lines). On the contrary, when the pump electric field was directed along either one of the two crystallographic axes, only Raman phonons were detected (Fig. 3d, blue line). Hence the magnetic response could only be driven when two different orthogonal phonon modes were excited. Moreover, the phase of the magnon oscillations switched sign when the pump polarization was rotated from +45 degrees to − 45 degrees, revealing a dependence on the relative phase of the two driven phonons (Fig. 3d, red and orange lines).

Finally, the phase of the measured magnon oscillations did not change when the static magnetization was reversed (Fig. 3e), indicating that their direction was independent from the initial canting of the spins.

Let us consider the physical situation depicted in Figure 1d, where the oxygen ions perform rotational motions. Qualitatively, these rotations are expected to modify the crystal field felt by the high-spin $Fe^{3+}$ ion at the centre of the octahedron. This effect mixes the ground state $t_{2g}^3 e_g^2$ electronic wavefunction[18] at each $Fe^{3+}$ ion with excited states, for which spin-orbit coupling is enhanced. Thus, the moving ions promptly perturb the angular momentum of the $t_{2g}^3 e_g^2$, triggering a magnetic excitation. This effect has some analogy with the case of electronic Raman excitation of magnons[19,20,21], in which visible light fields mix the ground state wavefunction with that of excited state orbitals.

Our results can be described by considering an effective Hamiltonian of the form $H_{eff} = i\alpha_{abc} Q_{ua} Q_{ub}^* M_c$. In this expression, $\alpha_{abc}$ is the magneto-elastic susceptibility, antisymmetric over the first pair of indices $\alpha_{abc} = -\alpha_{bac}$, $Q_{ua}$ and $Q_{ub}^*$ are the phonon eigenvectors and $M_c$ is the static magnetization. Hence, the circularly polarized lattice motion behaves as an effective magnetic field $\left[-\partial H_{eff}/\partial M_c\right] = -i\alpha_{abc} Q_{ua} Q_{ub}^*$ directed perpendicular to the rotation plane (the *ab* plane in the present case, Fig. 1d). Moreover, the sign of the field depends on the rotation direction of the ions.

These arguments, and especially the directionality of the effective field as perpendicular to the rotation plane of the atoms, are validated by temperature dependent measurements (figure 4).

When cooling below 96 K, ErFeO$_3$ undergoes a *spin reorientation transition* (SRT)[22], with a continuous rotation of the spins direction with respect to the crystallographic axes (Fig. 4a) and a progressive alignment of the total ferromagnetic moment with the *a* axis. Thus, for T<T$_{SRT}$, the *c*-axis-oriented effective magnetic field associated with the atomic loops is no longer parallel to the ferromagnetic moment (S$_1$+S$_2$), but becomes parallel to the antiferromagnetic moment (S$_1$-S$_2$). As a consequence[23,24], a different quasi-ferromagnetic (q-FM) mode is excited[17] (Fig. 4b), with a frequency that strongly depends on temperature (Fig. 4c)[22].

The effects above are further clarified when compared to calculations of the lattice and spin dynamics in ErFeO$_3$. We concentrate on the measurements taken at 100 K (T>T$_{SRT}$) and on the excitation of the quasi-AFM mode reported in figures 2 and 3. In the calculations reported in figure 5, we solved Maxwell's equations in time and space (see Supplementary Section 7), simulating the lattice polarization along each axis as mixture of driven Lorentz oscillators (Fig. 5a and 5b). The degree of rotation of the total polarization could then be estimated as $\left(\vec{P}_{ua+ub} \times \partial \vec{P}_{ua+ub}/\partial t\right)$ (see Supplementary Section 8). As displayed in the calculations of Figure 5c (light purple), a pulse of elliptical motion develops at the surface of the material immediately after the phonons are being excited with linear polarization. This dynamical lattice rotation "propagates" into the material at a speed of 0.4x10$^8$ m/s (13% c). This combination of two phonon-polaritons carries with it an effective magnetic field (Fig. 5d, see Supplementary Section 8) that results in the excitation of spin waves. Note that after a picosecond the rotation inverts its sign, although damping reduces the amplitude of the loops, causing a total breaking of time reversal invariance.

The magnon amplitude can be estimated at all points in time and space (Fig. 5e) by solving the Landau-Lifshitz equation[25], driven by the calculated time- and space-dependent effective field of Fig. 5d. The phase-matched signal measured by the probe (dotted lines in Fig. 5e) is displayed in the inset of Fig. 5e. By comparing the size of the calculated signal with the measured one, we can recover the total amplitude of the magnetic precession and the effective magnetic field strength, which is estimated to be 36 mT for a fluence of 20 mJ/cm$^2$.

Note that the effect reported here could be made larger by increases in the field strength, but also optimized by an appropriate choice of phonon resonances and by shaping the optical field to enhance phase matching. As the effect operates at mid-infrared and THz frequencies, it may become applicable in new devices that operate in this wavelength regime.

Finally, beyond the applications to magnetism discussed here, we note that control of ionic loops can be viewed as a perturbation of the Berry connection[26,27], and in appropriate circumstances may be used to manipulate the topological properties of materials[28, 29] through phonons.

**Acknowledgements**

We are grateful to A. Subedi, D. M. Juraschek, M. Fechner and N. A. Spaldin for sharing the calculated phonon eigenvectors and for useful discussions. We thank R.V. Pisarev for providing the samples. We additionally acknowledge support from J.Harms (for graphics), D. Nicoletti (for FTIR measurements) and G. Meier (for sample characterization).

A.V.K., R.V.M. and D.B. thank I. Razdolski for fruitful discussions of the first experiments and Th. Rasing for continuous support.

The research leading to these results received funding from the European Research Council under the European Union's Seventh Framework Programme (FP7/2007-2013)/ERC Grant Agreement no. 319286 (QMAC). We acknowledge support from the Deutsche Forschungsgemeinschaft via the excellence cluster 'The Hamburg Centre for Ultrafast Imaging - Structure, Dynamics and Control of Matter at the Atomic Scale'.

A.V.K., R.V.M. and D.B. acknowledge partial support by the European Union's Seventh Framework Program (FP7/2007-2013) Grant No. 281043 (Femtospin).

**Author contributions**

A.Cavalleri conceived the project together with T.F.N. T.F.N. and A. Cartella built the mid-IR pump/Faraday-rotation probe experimental set-up with help from A. Cantaluppi and M.F. T.F.N. and A. Cartella performed the measurements. A.Cartella and T.F.N. developed the 1D-FDTD code and performed the simulations with help from R.V.M. (for the solution of the LL equation). T.F.N. analyzed the experimental data. A.V.K., R.V.M. and D.B. identified the material system for the project and helped in the analysis of the data. R.M. collaborated in the interpretation of the data. The manuscript was written by T.F.N. and A. Cavalleri, with feedback from all co-authors.

**Competing financial interests**

The authors declare no competing financial interests.

**Data availability.** The data that support the plots within this paper and other findings of this study are available from the corresponding author upon reasonable request.

# FIGURES

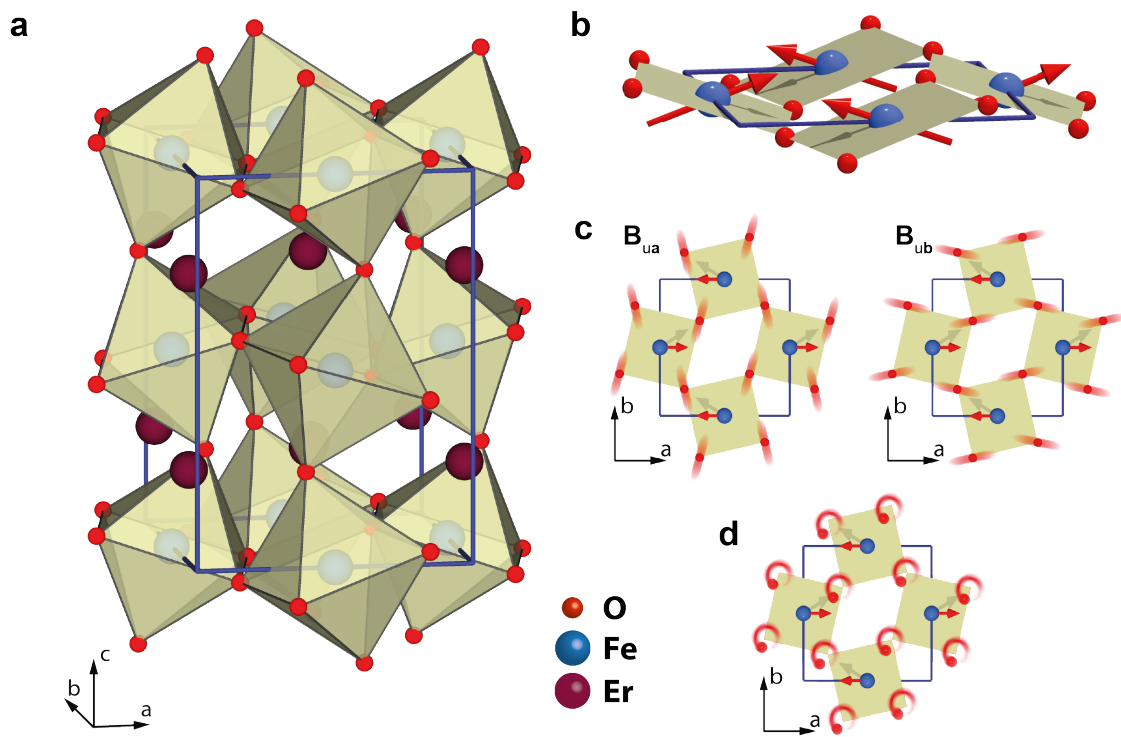

**Figure 1 | Structure and magnetic properties. (a) Crystal structure of ErFeO$_3$,** an orthorombically distorted perovskite (Pbnm). **(b) Magnetic ordering.** The spins of the iron ions order antiferromagnetically along *a*. Due to the Dzyialoshinkii-Moryia interaction, a canting is induced out of the *ab* plane resulting in a small ferromagnetic component along *c*. **(c) Eigenvectors of the highest frequency phonons excited by the pump pulse.** Singly degenerate infrared active B$_{ua}$ and B$_{ub}$ phonons (calculated eigenvectors for ErFeO$_3$). **(d) The motion of the ions results in an elliptically polarized phononic field due to the non-degenerate nature of the excited IR-active phonons (16.17 THz and 17.03 THz)**. All Fe-O layers in the unit cell behave in a similar fashion.

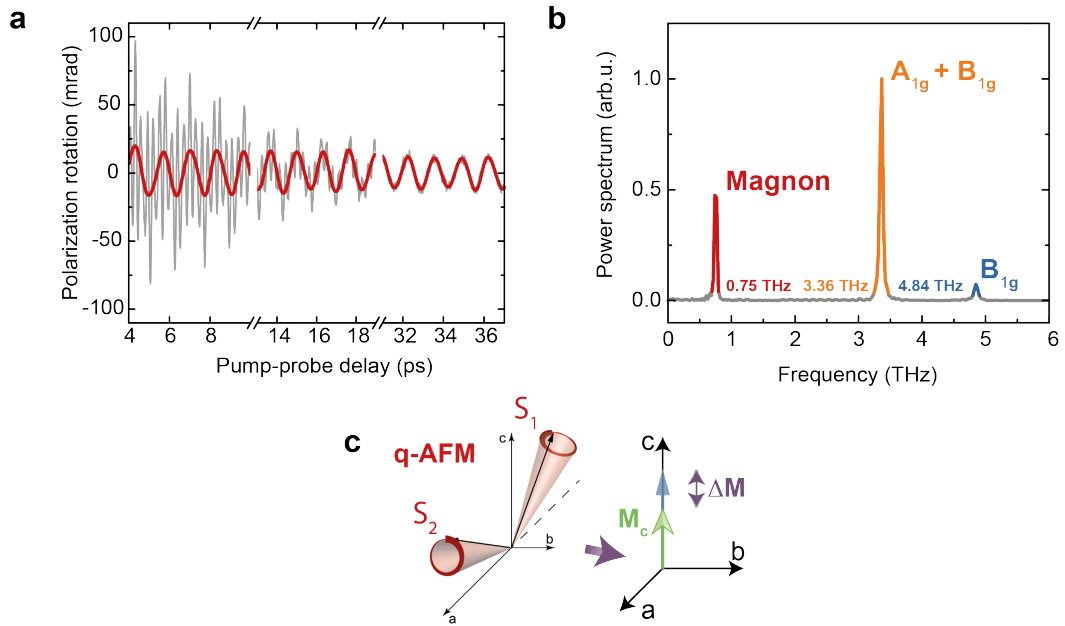

**Figure 2 | Transient birefringence measurement. (a) Pump-induced changes in the polarization of the probe as a function of pump-probe delay.** The slow-varying component has been subtracted. Multi-component fast oscillations (grey) can be filtered out by a low-pass filter (1.5 THz cut-off) to reveal the slow oscillation associated with the magnon (red). The sample was kept at 100 K while the fluence of the pump pulse was 17.6 mJ/cm$^2$. **(b) Power spectrum of the oscillatory signal.** The three peaks correspond to: 0.75 THz, quasi-antiferromagnetic (q-AFM) magnon (red); 3.36 THz and 4.85 THz, Raman phonons of symmetry $A_{1g}+B_{1g}$ (orange) and $B_{1g}$ (blue), respectively. **(c) Cartoon of the spin motion ($S_1$ and $S_2$) for the q-AFM mode.** The small ferromagnetic component along the *c* axis (green) oscillates in amplitude at the magnon frequency.

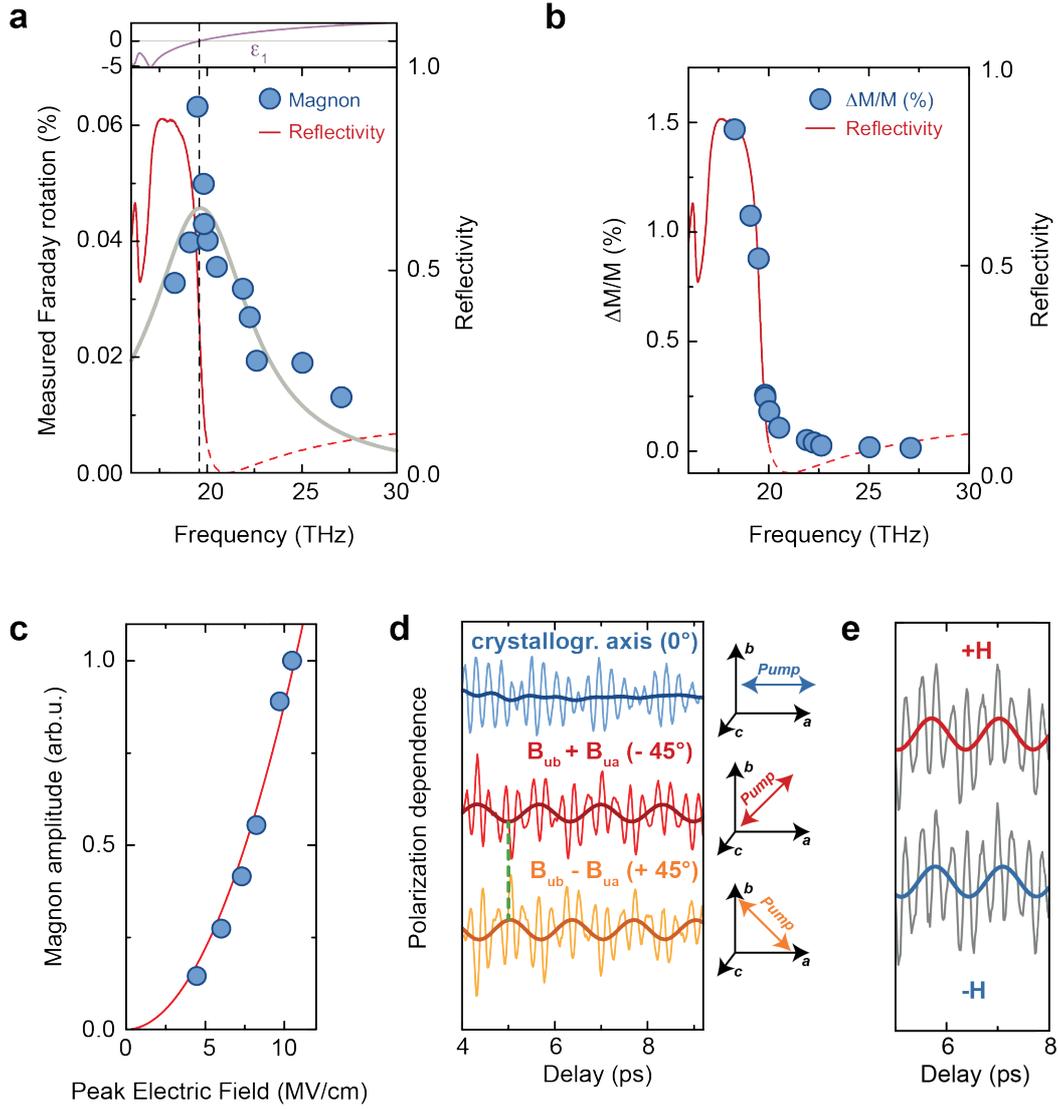

**Figure 3 | Pump wavelength, fluence, polarisation and external field dependence at 100 K. (a) Pump wavelength dependence of the measured Faraday rotation and real part of the dielectric function.** The blue dots are the values of the coherent magnon amplitude derived from a fit of the oscillations and extrapolated to zero time delay. The solid grey line is a lorentian fit intended as a guide to the eye. The solid red curve is the static sample reflectivity measured by Fourier transform infrared spectroscopy (FTIR). The dashed line is a fit to the measured reflectivity. The real part of the dielectric function $\varepsilon_1$ (solid purple line) is retrieved from the reflectivity via a Kramers-Krönig consistent fit. The dotted black line indicates the zero crossing of $\varepsilon_1$. All the mesurements were made at a constant fluence of 20 mJ/cm$^2$. **(b) Pump wavelength dependence of the magnetization change.** In order to estimate the absolute change of magnetization we corrected the data of Fig. 1a for the penetration depth mismatch between the pump and the probe (see Supplementary Section 6). **(c) Amplitude dependence on the pump field measured at 19.5 THz.** The magnon amplitude scales quadratically with the pump electric field - linear with the fluence. Indeed, the electric field ($E$) of a pulse can be estimated from its duration ($\Delta\tau$) and the fluence ($F$) at the sample position. The connecting formula is $E = \sqrt{(2F)/(\varepsilon_0 c \Delta\tau)}$. **(d) Pump polarization dependence.** Here are presented (light colours) the measured oscillations for different pump polarizations. The darker lines display the same data after application of a numerical low-pass filter (1.5 THz cut-off) to emphasise the magnon (when present). Upper curve: pump polarization directed along one crystallographic axis. Only the Raman-active phonon oscillations (*light blue*) can be detected while the magnon vanishes (dark *blue*). Middle curve: pump polarization in between *a* and *b*. The two IR-active phonons are excited with the same initial positive phase ($B_{ub} + B_{ua}$). In addition to the Raman phonons, the magnon appears (*dark red*). Lower curve: pump polarization rotated by 90 degrees, in between -*a* and *b*. The two IR-active phonons are excited with phases of opposite signs ($B_{ub} - B_{ua}$). The magnon experiences a π phase shift (*dark orange*). **(e) External magnetic field dependence.** The phase of the magnon does not depend on the initial orientation of the magnetic order.

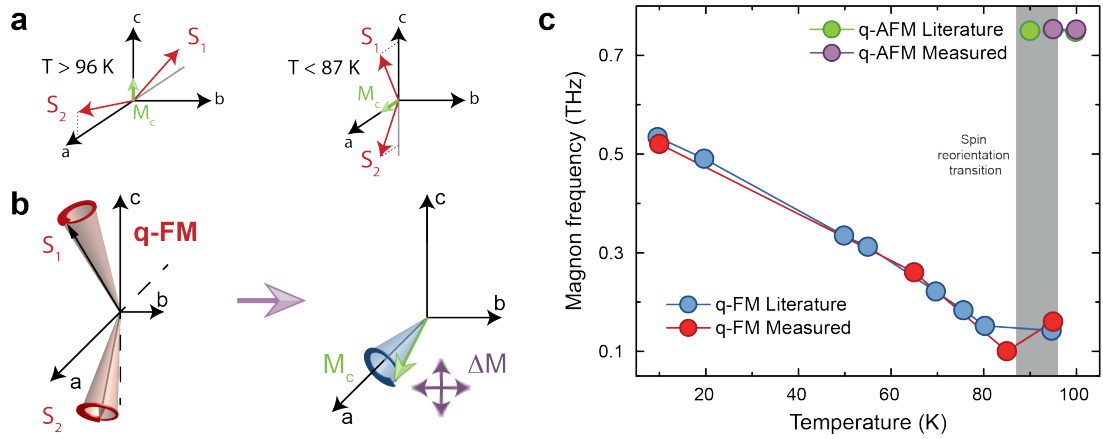

**Figure 4 | Temperature dependence. (a) Spin Reorientation Transition.** In ErFeO$_3$ for 87 K < T < 96 K a change of easy axis takes place. As a consequence the spins (red) and the resulting ferromagnetic moment (green) continuously rotate from the *c* axis to the *a* axis. The spin configuration varies from Γ$_4$(G$_x$F$_z$, magnetic space group: Pb'n'm[30]) for T>T$_{SRT}$ to Γ$_2$(G$_z$F$_x$, magnetic space group: Pbn'm'[30]) for temperatures below the spin reorientation transition. **(b) Cartoon of the spin motion for the quasi-ferromagnetic mode.** The small ferromagnetic component precesses around the *a* axis (light blue) resulting in a modulation of the magnetization along the *b* and *c* axes (purple arrows). **(c) Quasi-ferromagnetic (q-FM) magnon frequency temperature dependence.** Due to the spin reorientation transition the geometry of our experiment changes and we are able to excite exclusively the other magnon branch. The measured frequencies and the data from the literature (adapted from Ref [22]) are in excellent agreement. The time dependent traces for each temperature are shown in Supplementary Figure S5.

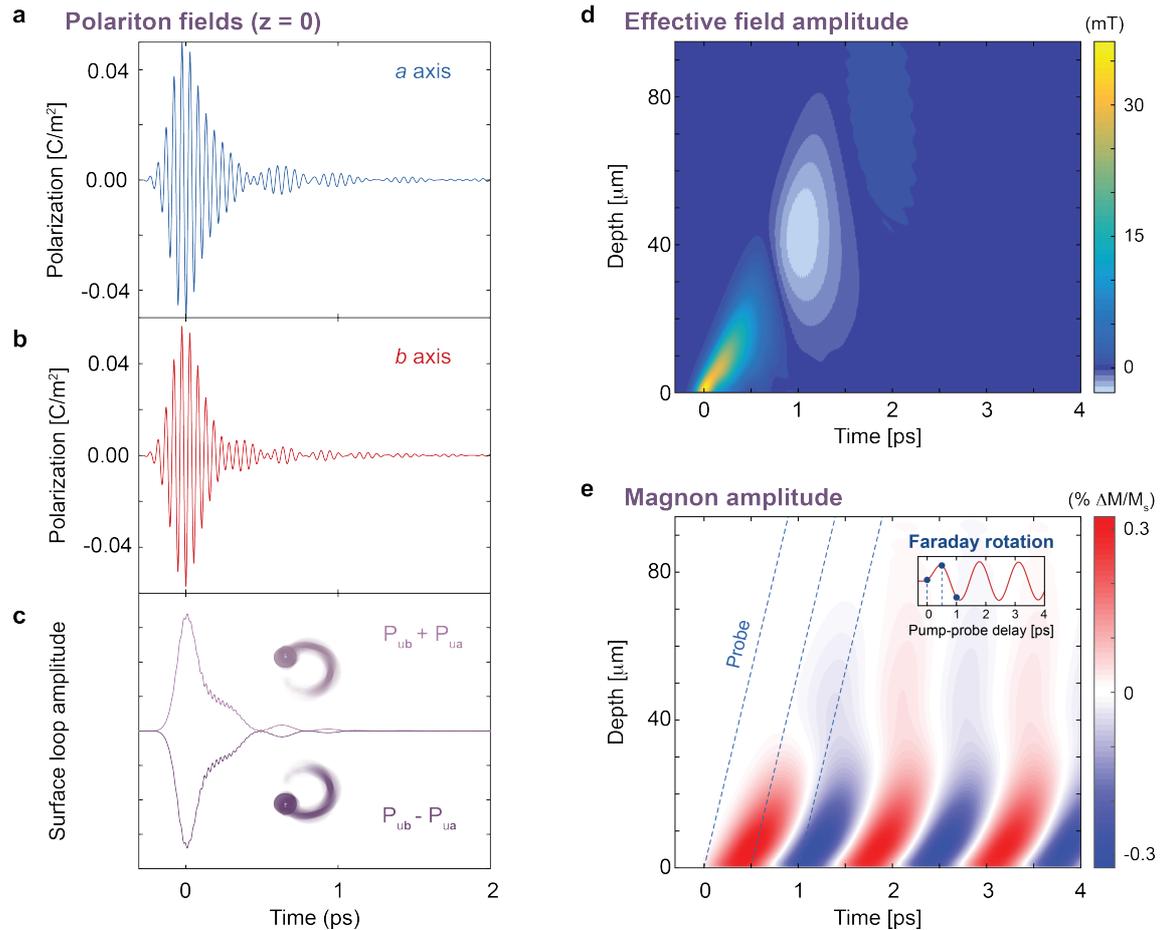

**Figure 5 | Calculated ionic and magnetic dynamics. (a), (b) Time-dependent polarisation at the surface of the material.** A 10 MV/cm amplitude, 130 fs long pulse tuned around 19.5 THz initiates the phonon-polariton response of the material along axes *a* and *b*, respectively. **(c) Circularity of the surface polarisation.** Due to the non-degenerate nature of the excited polar modes the ions on each axis react to the same driver with different initial phases and evolve at different rates. Hence, the total lattice polarization inside the material is elliptical. Furthermore, the initial phase of the IR-active phonons (light and dark purple) determines the initial direction of rotation of the ions and, consequently, the sign of the effective magnetic field. **(d) Elliptically polarized phonon-polariton propagation.** The pair of phonon-polaritons propagates inside the material at a speed of $0.4 \times 10^8$ m/s and carries an effective magnetic field. **(e) Solution of the Landau-Lifshitz (LL) equation for the q-AFM magnon.** We solved the LL equation in time for every layer of material (space resolution) considering as a driver the time-dependent phonon-generated atomic-loop amplitude of Fig. 5d. We then integrated only the signal seen by the propagating probe (oblique dotted blue line) for different pump-probe delays and normalized to the static magnetization value (inset Fig. 5e). A fitting parameter was used to match the simulated magnon amplitude with the measured one in order to estimate the value of the *effective magnetic field* generated by the phonons.

**References Main text**

**Methods**

**Experimental Setup**

The mid-IR pulses used in the experiment were generated by difference frequency generation (DFG) in a GaSe crystal between the signal outputs of two optical parametric amplifiers (OPAs), pumped with 5-mJ pulses, 100 fs long, at 1 KHz repetition rate and 800 nm wavelength delivered from a commercial regenerative amplifier. The two OPAs were seeded by the same white light continuum (WLC), producing phase locked[1] signal pulses (300uJ, ca. 70 fs long, independently tunable from 1.2 $\mu$m to 1.5 $\mu$m). As a consequence, the generated mid-IR transients were carrier-envelope-phase (CEP) stable (>6 $\mu$J across the 18 to 36 THz range, ca. 100fs long)[2]. The pump pulses were characterized by electro-optic sampling (Supplementary Section 2) and by a Michelson Fourier Transform Interferometer.

A small fraction of the regenerative amplifier output (100 fs long) was used to probe the Faraday rotation using a balanced detection configuration. A half waveplate and polarizer were put in front of the sample to align the polarization of the incoming light along one of its crystallographic axes (*a* or *b*). The angle between pump (normal incidence) and probe was 18 degrees. After the sample, a second waveplate and a Wollaston prism were used to balance the intensity on two identical diodes, whose differential signal was measured by a lock-in amplifier as a function of the pump-probe delay (an optical chopper was put on the Mid-IR beam).

The schematic representation of the setup is shown in Supplementary Figure S1.

**Sample**

The sample used in the experiment is a bulk single crystal, 95 micrometers thick, *c*-cut, grown using the floating zone method.

During the measurements the sample was kept in an external magnetic field, directed along the c axis, and strong enough to saturate the small ferromagnetic moment.

The orthoferrite temperature was controlled with a cold finger cryostat.

The in-plane axes were determined by static birefringence measurements (Supplementary Section 3). The polarization resolved reflectivity was obtained by means of Fourier Transform Infrared Spectroscopy (Supplementary Section 4).

**Simulations**

The dynamical properties of ErFeO$_3$ were obtained by solving Maxwell's equations in space (1D) and time for each axis (Supplementary Section 7). The lattice response was modeled by expressing $\varepsilon_r$ as a series of damped harmonic oscillators corresponding to the IR active phonons, plus an $\varepsilon_\infty$ taking into account the permittivity due to high energy excitations (Lorentz model):

$$\varepsilon_r = \varepsilon_\infty + \sum_k \frac{\omega_{p,k}^2}{\omega_{0,k}^2 - \omega^2 + i\omega\Gamma_k}$$

The values of the plasma frequency $\omega_p$, the TO frequency $\omega_0$ and the damping $\Gamma$ for each of the $k$ phonons were extracted from a Lorentz fit to the static reflectivity measurements.

The atomic displacement and the resulting effective magnetic field were estimated from the calculated lattice polarization (Supplementary Section 8).

The magnetic dynamics was obtained by solving the Landau-Lifshitz equation following the approach developed in Ref. 25 of the main text. The calculated effective magnetic field was used as the source term.

**References Methods**